\documentstyle[12pt]{article}
\begin{document}
\baselineskip=20pt
\pagestyle{myheadings}
\newcommand{\Real}{{\rm I\!R}}
\newcommand{\CP}{{\rm CI\!P}}
\newcommand{\Complex}{{\rm I\!\!\!C}}
\newcommand{\N}{\mbox{$\cal N$}}
\newcommand{\Sp}{\mbox{${\cal S}$}}
\newcommand{\U}{\mbox{$\cal U$}}
\newcommand{\Hi}{\mbox{$\cal H$}}
\newcommand{\Pe}{\mbox{$\cal P$}}
\newcommand{\Pc}{\mbox{${\cal P}^c$}}
\newcommand{\M}{\mbox{$\cal M$}}
\newcommand{\R}{\mbox{$\cal R$}}
\newcommand{\C}{\mbox{$\cal C$}}
\newcommand{\kb}{\mbox{\boldmath $\kappa$}}
\newcommand{\gb}{\mbox{\boldmath $\gamma$}}
\newcommand{\eb}{\mbox{\boldmath $e$}}
\newcommand{\Ab}{\bf A}
\newcommand{\Bb}{\bf B}
\newcommand{\Cb}{\bf C}
\newcommand{\Db}{\bf D}
\newcommand{\mub}{\mbox{\boldmath $\mu$}}
\newcommand{\muhatb}{\mbox{\boldmath ${\hat \mu}$}}
\newcommand{\Psib}{\mbox{\boldmath $\Psi$}}
\newcommand{\Phib}{\mbox{\boldmath $\Phi$}}
\newcommand{\Chib}{\mbox{\boldmath $\chi$}}
\newcommand{\T}{\bf T}
\def\i{\iota}
\def\e{\epsilon}
\def\Vstar{V^*}
\def\Vbar{\overline{V}}
\def\xibar{\overline{\xi}}
\def\t{\theta}
\def\p{\phi}
\def\k{\kappa}
\def\a{\alpha}
\def\b{\beta}
\def\Del{{\cal D}}
\noindent
\centerline{{{\large\bf The Geometric Phase }}}\\*[0.2cm]
\centerline{{{\large\bf and Ray Space Isometries}}}\\*[0.2cm]
\bigskip
\bigskip

\centerline{Joseph Samuel}
\centerline{Raman Research Institute}
\centerline{Bangalore, India 560 080}
\bigskip
\bigskip

\begin{abstract}
We study the behaviour of the geometric phase under isometries
of the
ray space. This leads to a better understanding of a theorem
first proved by Wigner: isometries of the ray space can always be
realised as projections of unitary or anti-unitary transformations
on the Hilbert space.
We suggest that the construction involved in Wigner's 
proof is best viewed as an use of the 
Pancharatnam connection to ``lift'' a ray space isometry 
to the Hilbert space.
\end{abstract}

\vspace*{6cm}

\vspace*{.5cm}\\
To appear in Pramana J. Phys.
\bigskip
\newpage
\section{Introduction}
The states of a quantum system are in one-to-one correspondence
with rays in Hilbert space. The ``overlap'' between rays is a measure of
the distance between them and can be directly 
measured in the laboratory as a transition
probability. A symmetry \cite{Fo1} of 
a quantum system maps 
the ray space onto itself 
preserving distances -- i.e, it is a ray
space isometry. Since it is inconvenient to work
directly on the ray space (defined as an equivalence class of 
states in Hilbert space), most quantum mechanical calculations are carried
out in Hilbert space.
Wigner \cite{Wigner} proved that any ray space
isometry can be realised on the Hilbert space of quantum mechanics
by either a unitary or antiunitary transformation. This theorem underlies
much of the study of symmetry in quantum mechanics. 
A complete and elementary account
of Wigner's proof of his theorem is given 
by Bargmann \cite{Bargmann}. 
More abstract and axiomatic accounts exist \cite{Emch}. 
Our purpose here, is to use geometric 
phase ideas \cite{Book}, which have recently been of interest,
to shed light on Bargmann's exposition of Wigner's theorem. This work follows
on an observation by Mukunda and Simon \cite{Mukunda1} regarding the
relation between Bargmann's paper and the geometric phase. 
This paper is structured as follows. 
We first review some well known facts
about the geometric phase to set this paper in context and 
then fix our notation in section 2. In section 3, we study 
the behaviour of the Pancharatnam excess phase under ray space isometries.
In section 4 we state Wigner's theorem and discuss its
significance. In section 5 we use the result of section 3 to prove
Wigner's theorem by an explicit construction. Section 6 is a concluding
discussion.  A fine point from section 3 is relegated to an appendix.

Berry \cite{Berry} noticed many years ago that standard treatments of the
adiabatic theorem in quantum mechanics had overlooked an
important phenomenon: when a quantum system in a slowly changing
environment returns to its original ray, the 
state of the system picks up an extra
phase of geometric origin above and beyond the phase
that one naively expects on dynamical grounds. Berry's phase attracted wide
attention because of its essentially geometric
character. The phase depends only on the path traversed by the system in ray
space and not on its rate of traversal. It is a measureable and gauge
invariant quantity, independent of phase conventions. 
An important paper by
Barry Simon \cite{Simon} shows that the Berry's phase is a
consequence of the curvature of the natural connection on a  
line bundle over the ray
space. Berry's original observation was made in the context of
the adiabatic theorem of quantum mechanics. Aharonov and Anandan
\cite{AA} showed how one could see Berry's phase even in nonadiabatic
situations. The key input here was to identify the dynamical
phase as the time integral of the 
expectation value of the Hamiltonian. When this dynamical
phase is removed, the geometrical picture described by Simon \cite{Simon}
applies. Although one starts with the Schr\"odinger equation,
after one identifies and removes the dynamical phase the
resulting parallel transport law is {\it purely} kinematic and
depends only on the geometry of Hilbert space. For this reason,
Berry's phase is also known as the ``geometric phase''.

It was pointed out by Ramaseshan and Nityananda \cite{RSRN} 
that Berry's
phase had been anticipated by Pancharatnam \cite{Panch} 
in his studies of the
interference of polarised light in the fifties. They showed that
Berry's phase for a two state system was a special case of
Pancharatnam's general study of interference of polarised light.
Pancharatnam had given a physically motivated criterion for comparing
the phases of two beams of polarised light. He went on to notice
that this criterion was not integrable. Two beams {\bf A} and
{\bf B} in
phase with a third beam {\bf C} are not in phase with each other. The
phase difference between {\bf A} and {\bf B} is equal to half the solid
angle subtended by the triangle ({\bf ABC}) on the Poincar\'{e} sphere 
\cite{Poincare,RSGNR}.
In the limit that
the discrete points approach a continuous curve, the Pancharatnam
phase reduces to Berry's phase for a two state system.

In reference \cite{Gene} Pancharatnam's ideas were carried over to
the Hilbert Space of quantum mechanics. The Pancharatnam criterion 
was used to compare the phases of states on any two non-orthogonal rays.
One defines two states to be ``in phase'' if their inner product
is real and positive.
This permits us to define the Pancharatnam lift: Given a discrete sequence of 
rays (successive rays not orthogonal), one can use the Pancharatnam
connection to ``lift'' the discrete set of rays to Hilbert space. This 
connection contains the natural connection as a special case and tends to it
in the limit that the sequence of points becomes a continuous curve.

The importance of
geodesics on the ray space of quantum mechanics was emphasized
in 
Ref. \cite{Gene}, which states and proves the geodesic rule:
Pancharatnam's 
criterion is equivalent to parallel transport of the
phase along the shortest geodesic in the ray space.  Given three
non orthogonal rays, one finds that
Pancharatnam's excess phase
is the integral of a two form over a {\it geodesic} triangle in
the ray space. This is the direct analogue of Pancharatnam's
``half the solid angle'' result. This general
framework permits an extension of Berry's work to nonunitary and
noncyclic situations.
Needless to say, this work also subsumes
the unitary and cyclic situations as a special case. It is also
observed in Ref.\cite{Gene} that Berry's phase appears in
systems subject to quantum measurements.  Analogue optical experiments
demonstrating this effect are reported in Refs. \cite{Bhandari,Hari}.
A review of
the field and a collection of papers upto 1989 is contained
in the book by Shapere and Wilczek\cite {Book}. A more recent and 
detailed treatment is given by Mukunda
and Simon \cite{Mukunda1}, who note the 
connection between Pancharatnam's excess phase
and invariants considered by Bargmann \cite{Bargmann}. 
Here, we 
follow on this observation made in \cite{Mukunda1}.
We show how Pancharatnam's
connection can be used to better understand a construction due
to Wigner. We show below how
one can use the Pancharatnam connection to
``lift'' a given ray space isometry to the Hilbert space.
\section{Preliminaries}
Let $\Hi$ be the Hilbert space of a quantum system and $\N:= \Hi
-\{0\}$ the space of normalisable states. We define \cite{Fo2} 
rays to be equivalence classes of normalisable states differing only
by multiplication by a nonzero complex number. We define two
elements  $|\Psi_1>$ and $|\Psi_2>$ of
$\N$ to be equivalent
( $|\Psi_1> \sim |\Psi_2>$)
if $|\Psi_1> =\a|\Psi_2>$, where $\a \in \Complex$,$\a\neq
0$. The ray space is defined as the quotient of $\N$ by this
equivalence relation.
$$\R=\N/\sim.$$
Elements of both $\Hi$ and $\N$ will be written as kets $|\,\,>$. 
 The natural projection
$$\Pi: \N \rightarrow \R$$
maps each normalizable state $|\Psi>$ to the ray $\bf \Psi$
on which it lies. We define the {\it overlap} between two rays
$\Psib_1$ and $\Psib_2$ as follows:
$$|\Psib_1.\Psib_2|^2:= \frac{<\Psi_1|\Psi_2><\Psi_2|\Psi_1>}
{<\Psi_1|\Psi_1><\Psi_2|\Psi_2>}.$$
By Schwartz inequality, $|\Psib_1.\Psib_2|\leq 1$ and $|\Psib_1.\Psib_2|=1$
if and only if $\Psib_1=\Psib_2$.
We define the distance $\delta(\Psib_1,\Psib_2)$ between the
rays $\Psib_1$ and $\Psib_2$ by 
$$|\Psib_1.\Psib_2|=\cos(\delta/2),$$
where $\delta$ lies between zero and $\pi$. Note that
$\delta(\Psib_1,\Psib_2)=0$ if and only if 
$\Psib_1=\Psib_2$.

Let $\{\gb(\lambda), 0\leq\lambda\leq 1\}$ be a curve in $\R$ and $|\gamma(0)>$
a vector on the ray $\gb(0)$. We define the ``horizontal lift''
of $\gb(\lambda)$ as the unique curve $|\gamma(\lambda)>$
starting from $|\gamma(0)>$ which satisfies
$\Pi(|\gamma(\lambda)>=\gb(\lambda)$ and 
\begin{equation}
<\gamma(\lambda)|\frac{d \gamma}{d \lambda}>=0
\label{hor}
\end{equation} 
Equation (\ref{hor}) gives us a rule (mathematically a
connection) for comparing vectors on neighbouring
rays. 

The Pancharatnam connection which we now describe  is a more
general notion that permits a comparison of vectors on {\it any} two non-
orthogonal rays.
Let $|A>$ and $|B>$ be two non-orthogonal vectors. We define
them as being ``in phase'' if the inner product
$<A|B>$ is real and positive. Given $|A>$ and a
ray $\Bb$, there is an unique $|B>$ which is {\it in phase}
with $|A>$ and has the same size
($<B|B>=<A|A>$). We refer to $|B>$
as the Pancharatnam lift of $\Bb$ (with $|A>$ as reference).

Given three pairwise non-orthogonal rays ${\bf A},\Bb,{\bf C}$, one can define the quantity
\begin{equation}
\Delta_{{\Ab}{\Bb}{\Cb}}=\frac{<A|B><B|C><C|A>}
{<A|A><B|B><C|C>}
\label{Bargmann}
\end{equation}
where, $|A>,|B>$ and $|C>$ are representative 
elements from the corresponding rays. $\Delta_{{\Ab}{\Bb}{\Cb}}$ depends
only on the rays ${\bf A},\Bb,{\bf C}$ and not on the representatives.
We will sometimes abbreviate $\Delta_{{\Ab}{\Bb}{\Cb}}$ to $\Delta$.
The phase $\beta$ of the complex
number $\Delta=\rho \exp(i\beta)$ is the Pancharatnam excess phase,
which is well defined (modulo $2 \pi$) if $\rho\neq0$.

An {\it Isometry} of the ray space is a map
\begin{equation}
{\bf T}:\R\rightarrow\R
\label{isometry}
\end{equation}
which preserves distances. Writing 
\begin{equation}
\Psib'={\bf T} \Psib,
\end{equation}
${\bf T}$ is an isometry if
\begin{equation}
|{\bf A}^{'}.\Bb^{'}|=|{\bf A}.\Bb|.
\label{iso}
\end{equation}
Under isometries the rays ${\Ab},{\Bb},{\Cb}$ go to ${\Ab}',{\Bb}',{\Cb}'$
and $\Delta_{{\Ab}{\Bb}{\Cb}}$ goes to $\Delta_{{\Ab}'{\Bb}'{\Cb}'}$, which
we will abbreviate to $\Delta'$.
\section{Isometries and the Pancharatnam Phase}
We now study the transformation of $\Delta$ under ray space isometries.
Let ${\bf A},\Bb,{\bf C}$ be three distinct pairwise 
non-orthogonal rays. Let us choose
unit representatives $|A>,|B>,|C>$ from these rays. Further,
let us choose the phases of these representatives so that $|B>$ 
is in phase
with $|A>$ (their inner product  $<A|B>$ is real and positive)
\begin{equation}
<A|B>=\cos c/2
\label{cosc}
\end{equation}
and 
$|C>$ 
is in phase
with $|A>$
\begin{equation}
<C|A>=\cos b/2.
\label{cosb}
\end{equation}
This of course means that $|C>$ is not (in general) in 
phase with $|B>$. In fact,
\begin{equation}
<B|C>=\cos a/2 \exp(i\beta).
\label{beta}
\end{equation}
$a,b$ and $c$ above are the distances (lengths
of the shortest geodesics in $\R$) between the rays
$({\Ab},{\Bb},{\Cb})$.  $(a,b,c)$ are the sides of the geodesic triangle
with vertices $({\Ab},{\Bb},{\Cb})$ and take values strictly between 
$0$ and $\pi$.

Let 
\begin{equation}
|\mu_B>=|B>-\cos(c/2) |A>
\label{mu2}
\end{equation}
be the component of $|B>$ orthogonal to $|A>$. 
Since 
$<\mu_B|\mu_B>=\sin^2(c/2)$,
we define the unit vector 
\begin{equation}
|{\hat \mu}_B>=|\mu_B>/\sin(c/2).
\label{mu2hat}
\end{equation}
Using $|A>$ and $|{\hat \mu}_B>$ as an orthonormal basis in
the $|A>-|B>$ plane, one sees (on the Poincar\'{e} 
sphere) that the horizontal curve $ \{|\gamma_B(\lambda)>, 0\leq\lambda\leq 1\}$,
joining $|A>$ to $|B>$
($|\gamma_B(0)>=|A>, |\gamma_B(1)>=|B>)$
\begin{equation}
|\gamma_B(\lambda)>=\cos(\lambda c/2) |A>+\sin(\lambda c/2) 
|{\hat \mu}_B>
\label{C2}
\end{equation}
projects down to the shortest geodesic $\gb_B(\lambda)$
connecting ${\bf A}$ and $\Bb$ ($\gb_B(0)={\bf A}, {\gb}_B(1)=\Bb$).

The tangent vector to the curve $|\gamma_B(\lambda)>$ at $\lambda=0$ is
\begin{equation}
|{\dot \gamma_B(0)}>= (c/2)|{\hat \mu}_B>
\label{C2dot}
\end{equation}
Similarly $ \{|\gamma_C(\lambda)>, 0\leq\lambda\leq 1\}$ defined
as
\begin{equation}
|\gamma_C(\lambda)>=\cos(\lambda b/2) |A>+\sin(\lambda b/2) |{\hat \mu}_C>
\label{C3}
\end{equation}
is the horizontal lift of the shortest geodesic 
connecting ${\bf A}$ with ${\bf C}$. In (\ref{C3}) 
$|{\hat \mu}_C>$ is
the normalised vector $|{\hat \mu}_C>=|\mu_C>/\sin(b/2)$ where
$|\mu_C>=|C>-\cos(b/2) |A>$.
The tangent vector to the curve $|\gamma_C(\lambda)>$ at $\lambda=0$ is 
\begin{equation}
|{\dot \gamma_C(0)}>= b/2 |{\hat \mu}_C>.
\label{C3dot}
\end{equation}
The angle $A$ between the 
geodesics $\gb_B(\lambda)$ and $\gb_C(\lambda)$ at ${\bf A}$
is given by 
\begin{equation}
\cos (A)=\frac{\Re(<{\dot \gamma}_B|{\dot \gamma}_C>)}{(<{\dot \gamma}_B|{\dot
\gamma}_B><{\dot \gamma}_C|{\dot \gamma}_C>)^{1/2}},
\end{equation}
where $\Re(\alpha)$ means the real part of $\alpha$.
This is easily worked out as 
\begin{equation}
\cos (A)=\frac{\cos(a/2)\cos (\beta)-\cos(b/2) \cos (c/2)}{\sin(c/2) \sin(b/2)}.
\label{Cosa}
\end{equation}
This gives us the formula 
\begin{equation}
\cos (\beta)=\frac{\cos(A)\sin(c/2)\sin(b/2) +\cos(b/2) \cos (c/2)}{\cos(a/2)}
\label{Major}
\end{equation}
for the cosine of the Pancharatnam phase.
The right hand side of this equation contains only the sides $a,b,c$ and
(one of) the
angles
of the geodesic triangle connecting the rays ${\Ab},{\Bb},{\Cb}$. All these
quantities are 
manifestly invariant under isometries of the ray space. It follows that 
$\cos(\beta)$ is also an isometry invariant. Since $\rho=|\Delta|$ is clearly
isometry invariant, it follows that $\Re(\Delta)$ is isometry invariant and
hence that 
\begin{equation}
\Delta'=\chi(\Delta)
\label{delta}
\end{equation}
 where $\chi(\alpha)=\alpha$
or $\chi(\alpha)={\overline \alpha}$. (\ref{delta}) is valid for {\it all}
triplets of
rays (including orthogonal ones, for which it becomes trivial).
Since the map $\T$ is continuous,
the function $\chi$ must be the {\it same} all over the ray
space (see appendix) and
can be determined \cite{Bargmann} from $\T$ \cite{Fo3}.

\section{Statement of the Theorem}
We address the following problem. Given a ray space isometry ${\bf T}$,
construct a map $T:\N\rightarrow\N$ so that the following diagram commutes
\begin{eqnarray*}
\N&\stackrel{T}{\longrightarrow}&\N\\
 {}^{\Pi}\downarrow &           &\downarrow{}^{\Pi}\\
\R&\stackrel{\T}{\longrightarrow}&\R\\
\end{eqnarray*}
or algebraically,
\begin{equation}
{\Pi(T(\Psi>))}={\T}(\Pi(|\Psi>))
\end{equation}
$T$ is called the ``lift'' of ${\bf T}.$
Clearly, there are many such lifts $T$ since, given $|\Psi>$, we
could pick as its image $|\Psi'>$ an arbitrary 
point from the fibre above ${\T}(\Pi(|\Psi>))$. 
We could in fact turn this nonuniqueness to advantage 
and demand that the lift $T$ has 
some nice properties. For instance, we could demand
that $T$ be continuous. We {\it will} assume below that $T$ is continuous
but even this restriction allows much
residual freedom. For example,
if $\T$ is the identity map, for each 
continuous, nonzero complex  function $f$ on
$\N$,  $T_f$ defined by  $T_f(|\Psi>)=f |\Psi>$ is a continuous lift.
Clearly, we can do much better and demand that $T$ has some
{\it more} nice properties. The conditions we impose should be
as strong as we can demand (so 
that the lift has desirable properties and is reasonably unique)
and yet weak enough
that a lift exists.
Continuity of $T$ is clearly too weak.
We are free to impose more conditions on the lift $T$. Wigner's theorem
does just that. Wigner showed that one can find a continuous lift which
preserves intensities ({\bf W1} below) as well as superpositions
({\bf W2} below).

{\it Wigner's theorem:}
There exists a lift $T$ of $\T$ which
\begin{description}
\item [W1] satisfies $<\Psi'|\Psi'>=<\Psi|\Psi>$
\item [W2] when extended to $\Hi$ by $T|0>=|0>$ satisfies
$$T(|A>+|B>)=T(|A>+|B>)$$
\end{description}
The lift is unique upto an overall phase \cite{Fo3}.

The content of Wigner's theorem is that {\it all} ray space
isometries (i.e all maps $\T$
which satisfy (\ref{iso})) can be realised by maps on $\Hi$ satisfying
({\bf W1},{\bf W2}). No other isometries exist and nothing is
lost by restricting
attention to maps $T$ which satisfy ({\bf W1,W2}). We prove
Wigner's theorem below by explicitly constructing the map $T$.

\section{Wigner's Construction}
Let $|e>$ be any fixed vector in $\N$, $\eb$ its ray and $\eb'$  the image
of $\eb$ under $\T$. Let us arbitrarily pick $|e'>$ from $\eb'$ satisfying
$<e'|e'>=<e|e>$ and define $T|e>$ to be $|e'>$.
$|e'>$ is arbitrary up to a phase. This is the only
arbitrariness in the entire contruction which follows. 
Let $\Pe=\{|\Psi>\in \Hi|<e|\Psi>=0\}$ be the set of elements in $\Hi$
orthogonal to $|e>$. And let $\Pc$ be its complement-- the set of
elements in $\Hi$ which are {\it not} orthogonal to $|e>$.
We now define the action of $T$ on all elements of $\Pc$ using
the Pancharatnam lift.
Let $|\Psi>\in \Pc$ be such an element. From (\ref{iso}), it  follows
that $|(\Psib',\eb')|$ is not zero. We map $|\Psi>$ to the unique
element $|\Psi'>\in\Psib'$ which satisfies (\ref{amp},\ref{phase})
below.
\begin{equation}
<\Psi'|\Psi'>=<\Psi|\Psi>
\label{amp}
\end{equation}
determines the amplitude of $|\Psi'>$. Since $|<e'|\Psi'>|=|<e|\Psi>|$,
we can choose the phase of $|\Psi'>$ to satisfy
\begin{equation}
<e'|\Psi'>=\chi(<e|\Psi>).
\label{phase}
\end{equation}
It follows from (\ref{delta})
rewritten here as 
\begin{equation}
\frac{<e'|A'><A'|B'><B'|e'>}{<e'|e'><A'|A'>
<B'|B'>}=
\chi(\frac{<e|A><A|B><B|e>}{<e|e><A|A>
<B|B>})
\end{equation}
that if $|A>$ and $ |B>$
are any two vectors in $\Pc$,  $|A'>$ and $ |B'>$
defined as in (\ref{amp},\ref{phase}) above satisfy
\begin{equation}
<A'|B'>=\chi(<A|B>).
\label{Panch}
\end{equation}
Note that this lift preserves superpositions. For if $|\Psi>=
|A>+|B>$, (all $|\,\,>$ s in $\Pc$), 
a simple calculation shows that the norm 
of
$$|\phi'>=|\Psi'>-(|A'>+|B'>)$$
vanishes. It follows that 
\begin{equation}
|\Psi'>=|A'>+|B'>.
\end{equation}
Actually, more is true. If $|A>+|B>=|C>+|D>$
(all $|\,\,>$ s in $\Pc$),  we find that 
$|A'>+|B'>=|C'>+|D'>$.
The proof as before, is to just compute the norm of the
difference of both sides and use (\ref{Panch}).
Note that the sum $|A>+|B>$ need not be in $\Pc$. We
can therefore define the action of $T$ on elements of $\Pe$ by superposition.
Any element $|\Phi>\in \Pe$ can be written as
sums of elements in $\Pc$.
For example
\begin{equation}
|\Phi>=(|\Phi>-|e>)+|e>
\end{equation}
In fact there are many ways to express $|\Phi>$ as sums of elements
of $\Pc$. It doesn't matter which of these
ways one chooses and that the extension of $T$ to $\Pe$ is 
well defined. We have thus defined $T$ on all of $\Hi$ satisfying
({\bf W1,W2}).

\section{Conclusion}
 The key new observation of 
\cite{Mukunda1} which led to the present work 
is that the quantity $\Delta$ which has recently been of
interest in the context of the Pancharatnam phase is exactly
what was used by
Bargmann to discriminate between unitary and anti-unitary transformations.
Bargmann remarks
\cite{Bargmann} that one can determine the function
$\chi(\alpha)$ merely from a knowledge of the map $\T$ (for $dim
(\Hi)> 1$ \cite{Fo3}). One starts with $\Delta$, 
which is defined on the ray space. Using $\T$, one
determines $\Delta'$
and from $\Delta'=\chi(\Delta)$,
one can determine $\chi$.

The main difference between our exposition and Ref.\cite{Bargmann}
is that Bargmann deduces (\ref{delta}) as a {\it corollary},
after constructing a lift of ${\bf T}$. We reverse the order and,
using geometric phase ideas, first
prove (\ref{delta}) as a geometric identity on the ray space.
This result is then used as an {\it input} for constructing the
lift and showing that it does have the desired properties ({\bf W1,W2}).
This leads to a considerably simplified and elementary exposition of 
Wigner's theorem based on ideas from the geometric phase.

We have derived a formula (\ref{Major}) 
expressing the cosine of the Pancharatnam excess
phase in terms of isometry invariants. This leads to two
distinct possibilities for
the transformation of the Pancharatnam phase under isometries:
it is either preserved or reversed. The lift $T$ is accordingly unitary
or anti-unitary. Note that the Pancharatnam phase $\beta$ itself is {\it not}
an isometry invariant, but only its cosine. 
The non invariance of $\beta$ is {\it precisely}
what Bargmann uses to distinguish between unitary and
antiunitary transformations.

It is interesting 
to note that trigonometry in ray space is qualitatively different
from plane or spherical trigonometry. In ray space, 
the sides of a triangle $(a,b,c)$ do not determine
its angles $(A,B,C)$. To see this, it is enough to consider a $3$
(complex) dimensional Hilbert space $\Hi$ 
(since three rays are involved). A triangle
in $\R$ is determined by $3$ distinct rays in $\R$. Since $\R=\CP^2$ is
$4$ (real) dimensional, the set of triangles is $12$ (real) dimensional.
The isometry group of $\CP^2$ is $8$ (real) dimensional and acts
freely on triangles. It
follows that a triangle in the ray space has $4$ independent
isometry invariants.
We chose to  
express (\ref{Major}) $\cos(\beta)$ in terms of the four
independent variables $(a,b,c,A)$.
One could equally well choose any four of these six variables.

For simplicity, we assumed that a symmetry maps the ray space $\R$ to itself.
More
generally, one can have maps between different ray spaces. Such
a situation arises if there is more than one superselection
sector in the theory.
An example of such a mapping is charge conjugation, 
which maps different charge superselection sectors to each other.
Our analysis is easily adapted to mappings between different
superposition sectors. 

To mathematicians, the 
ray space is a K\"ahler manifold \cite{Arnold,Abbati}, 
with three interlinked structures:
a metric, a symplectic structure and a complex structure. Any
two of these determine the third. Physically, the metric represents
transition probabilities and the symplectic
2-form is the curvature of the natural connection
that emerges from Berry's phase \cite{Cantoni}. 
Isometries of $\R$ preserve the metric, but may reverse the symplectic
structure. This then means that the complex structure is also reversed.

We feel that this paper  provides an interesting application of
the Pancharatnam connection.
Note that the Pancharatnam connection has been used in an essential
way. The natural connection only permits a comparison of
neighbouring rays and therefore could be used only in the tangent
space around $|e>$. 
The global nature of the Pancharatnam
connection allows us to define a lift of $\T$ for (almost)
all rays at once. The gaps are then filled in by superposition.

{\it Acknowledgements:} The author thanks Rajaram Nityandanda for
several discussions and a critical reading of the manuscript.
\section*{Appendix}
In this appendix we  
show that continuity implies that $\chi$ is the same all
over ray space. There is a subtlety here stemming from the fact that there
are regions in ray space where $\Im(\Delta)$, the imaginary part 
of $\Delta$ vanishes and the two possibilities for $\chi$ coincide.
Let us fix rays ${\Ab},{\Bb}$
and consider $\Delta$ as a function of ray ${\Cb}$. Let us define $\R^+$
as the set of 
points of $\R$ where $\Im(\Delta)>0$ and similarly $\R^-$ is the set where
$\Im(\Delta)<0$. 
We first show that
$\R^+$ is path connected. 
 Let ${\Cb}$ and ${\tilde {\Cb}}$ be two rays in $\R^+$. Let us choose 
a representative vector
$|C>$ and 
decompose it into components  $|C^{\parallel}>$ in the 
$|A>-|B>$ plane and $|C^{\perp}>$  orthogonal
to it. By continuously decreasing the orthogonal component of
$|C>$ to zero, one can deform $|C>$ to
$|C^{\parallel}>$ in the $|A>-|B>$ plane. 
In the expression (\ref{Bargmann}) for $\Delta$,
$|C^{\perp}>$  does not contribute to the numerator and the
denominator is real and positive.
It follows that the sign of $\Im(\Delta)$ does not change in
the process of decreasing the orthogonal component of $|C>$ 
and so the deformation is entirely within $\R^+$. 
 Likewise $|{\tilde C}>$ can
also be deformed within $\R^+$ to $|{\tilde C}^{\parallel}>$ in 
the $|A>-|B>$ plane. The resulting kets are now in the two dimensional
subspace spanned
by $|A>$ and $|B>$ and 
we can now visualise the situation on the Poincar\'{e}
sphere.  Let ${\mbox{$\cal C$}}$
be the great circle through the points ${\Ab}$ and ${\Bb}$ on the
Poincar\'{e} sphere. ${\mbox{$\cal C$}}$ 
divides the sphere into two hemispheres. $\Im(\Delta)$ vanishes
only for points belonging to $\C$ and $\Im(\Delta)$ is strictly positive
on one hemisphere and strictly negative on the other. 
Since ${\Cb}$ and ${\tilde {\Cb}}$ belong to $\R^+$, the 
rays ${\Cb}^{\parallel}$ and ${\tilde {\Cb}}^{\parallel}$
corresponding to
the vectors $|C^{\parallel}>$, $|{\tilde C}^{\parallel}>$ lie
in the same hemisphere. They can therefore be deformed into each
other without passing through the equator. Throughout this deformation,
$\Im(\Delta)$ is positive and it follows that 
$\R^+$ is connected. (An identical argument shows that $\R^-$ is
connected.)

Since $\R^+$ is connected, continuity of ${\bf T}$ implies that 
$\chi$ must be the same all over $\R^+$.
Likewise, $\chi$ must be the same all over $\R^-$. If $\chi$
were to differ between between $\R^+$ and $\R^-$, both $\R^+$ and $\R^-$
would be mapped to the same component ($\R^+$ or $\R^-$). This contradicts
the fact that the map $\T$ is onto. Therefore $\chi$ must be the
same all over $\R$.


\begin{thebibliography}{99}
\bibitem{Fo1}
A symmetry of a quantum system is also required to leave the  
Hamiltonian invariant. This is not relevant here as we will not 
be concerned with Hamiltonians in this paper.
\bibitem{Fo2}
The definition of a ray here differs from \cite{Gene,Bargmann,Mukunda1}, 
where only a phase and not the amplitude is ``modded out''.
The present definition is the usual one in the mathematical
literature.
\bibitem{Fo3}
The case dim $\Hi=1$ is exceptional. In this case $\Delta$ is real and
$\T$ does not determine $\chi$.  Either choice of $\chi$ is allowed.
As a result there is an additional,
discrete, two-fold ambiguity in the lift.
\end{thebibliography}
\end{document}